\documentclass[twocolumn,english,prb,preprintnumbers,eqsecnum,amsmath,amssymb]{revtex4}
\usepackage[T1]{fontenc}
\usepackage[latin9]{inputenc}
\usepackage{amstext}
\usepackage{amsmath}
\usepackage{amsfonts}
\usepackage{amssymb}
\usepackage{mathtools}
\usepackage{graphicx}
\usepackage{rotating}
\usepackage {bm}
\usepackage[section]{placeins}
\usepackage{color}
\usepackage{multirow}
\usepackage{cellspace}
\makeatletter
\@ifundefined{textcolor}{}
{%
 \definecolor{BLACK}{gray}{0}
 \definecolor{WHITE}{gray}{1}
 \definecolor{RED}{rgb}{1,0,0}
 \definecolor{GREEN}{rgb}{0,1,0}
 \definecolor{BLUE}{rgb}{0,0,1}
 \definecolor{CYAN}{cmyk}{1,0,0,0}
 \definecolor{MAGENTA}{cmyk}{0,1,0,0}
 \definecolor{YELLOW}{cmyk}{0,0,1,0}
}
\def\Im{{\text{Im}}\,}

\def\epsilonF{\epsilon_{\text{F}}}
\def\kF{k_{\text{F}}}
\def\vF{v_{\text{F}}}
\def\NF{N_{\text{F}}}

\def\Tball{T_{\text{ball}}}
\def\tausp{\tau_{\text{sp}}}
\def\be{\begin{equation}}
\def\ee{\end{equation}}
\def\bea{\begin{eqnarray}}
\def\eea{\end{eqnarray}}
\def\bse{\begin{subequations}}
\def\ese{\end{subequations}}

\usepackage{dcolumn}
\usepackage{bm}

\makeatother

\usepackage{babel}
\begin{document}

\preprint{}

\title{Generic non-Fermi-liquid behavior of the resistivity in magnets with ferromagnetic, helical, or skyrmionic order}

\author{T.R. Kirkpatrick$^{1}$ and D. Belitz$^{2}$}

\affiliation{$^{1}$Institute for Physical Science and Technology, University of Maryland, College Park, MD 20742\\
 $^{2}$Department of Physics, Institute of Theoretical Science, and Materials Science Institute, University of Oregon, Eugene, OR 97403}

\date{\today}
\begin{abstract}
The electrical resistivity of several relatively clean metallic ferromagnets, as well as the helimagnet MnSi, is commonly observed
to exhibit non-Fermi-liquid behavior at low temperatures. This behavior, which is found in both ordered and disordered phases,
and both near and away from the magnetic transition, remains a major unsolved problem. We derive and discuss three novel
mechanisms underlying such behavior that are based on electron scattering mediated by the exchange of (1) ferromagnons
or (2) skyrmionic fluctuations, both in conjunction with weak disorder, or (3) helimagnons in clean systems. Since the magnetic
transition in weakly disordered sytems is generically discontinuous, static droplets of the ordered phase can exist within the
disordered phase, making the mechanisms viable there as well. We compare our theoretical results with existing experimental
ones and suggest additional experiments.  

\end{abstract}

\pacs{}

\maketitle

\section{Introduction, and Results}
\label{sec:I}

\subsection{Non-Fermi-liquid behavior in magnets}
\label{subsec:I.A}

\begin{figure}[t]
\includegraphics[width=8cm]{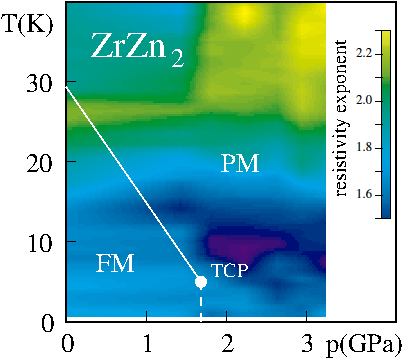}
\caption{Experimental temperature-pressure phase diagram of ZrZn$_2$ showing the ferromagnetic (FM) and paramagnetic (PM)
              phases. The tricritical point (TCP) separates the line of second-order transitions (solid) from the line of first-order
              transitions (dashed). The false colors indicate the exponent $s$ that characterizes the power-law $T$ dependence of the
              electrical resistivity $\delta\rho$. After Ref.~\onlinecite{Takashima_et_al_2007}.}
\label{fig:1}
\end{figure}

\begin{figure}[t]
\includegraphics[width=8cm]{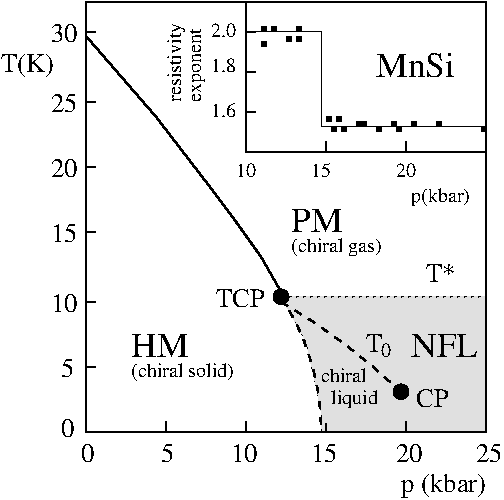}
\caption{Temperature-pressure phase diagram of MnSi combining experimental data from Refs.~\onlinecite{Pfleiderer_et_al_1997, Pfleiderer_2007,
              Pfleiderer_et_al_2004} with theoretical interpretations from Ref.~\onlinecite{Tewari_Belitz_Kirkpatrick_2006}. HM and PM are the helimagnetic
              and paramagnetic phases, respectively. The tricritical point (TCP) separates the line of second-order transitions (solid) from a line of
              first-order transitions (dashed). The shaded area is the NFL region; its upper limit (dotted line, $T^*$) is not sharp. The observed 
              resistivity exponent value changes abruptly at the critical pressure as shown in the inset. The region below $T_0$ is where
              Ref.~\onlinecite{Pfleiderer_et_al_2004} found partial helical order. This was interpreted in Ref.~\onlinecite{Tewari_Belitz_Kirkpatrick_2006}
              as a chiral liquid, separated from the chiral gas above $T_0$ by a first-order transition that ends in a critical point (CP). See the text for
              further information.}
\label{fig:2}
\end{figure}

A characteristic property of Fermi liquids is a low-temperature ($T$) electrical resisitivity $\rho$ of the form 
$\delta\rho(T\to 0) \propto T^2$, with $\delta\rho = \rho - \rho_0$ the $T$-dependent part of the resistivity and 
$\rho_0$ the residual resistivity;\cite{Abrikosov_Gorkov_Dzyaloshinski_1963} this behavior is realized in simple 
metals.\cite{Wilson_1954, Kittel_2005} A resistivity that does not satisfy this law is one of the hallmarks of metals 
referred to as non-Fermi liquids (NFLs). An obvious cause for NFL behavior is the vicinity of a quantum critical point, where the critical dynamics can lead to an 
unusual time and temperature dependence of various correlation functions.\cite{Hertz_1976, Millis_1993} More puzzling are 
examples where the NFL behavior is observed in large regions of the phase diagram, even far from any phase transition. 
One such class of NFL materials is comprised of metallic ferromagnets (FMs) and helimagnets (HMs). The NFL behavior can occur in either 
the magnetically ordered or disordered phases, or in both, and it is characterized by a $T^s$ behavior
\be
\delta\rho(T\to 0) = A_s T^s
\label{eq:1.1}
\ee
with an exponent $s$ that is commonly observed to be $3/2 \alt s < 2$. Examples include the ferromagnets ZrZn$_2$,\cite{Takashima_et_al_2007} Ni$_3$Al tuned by either
doping with Pd \cite{Sato_1975} or pressure,\cite{Niklowitz_et_al_2005} and Ca$_x$Sr$_{1-x}$RuO$_3$,\cite{Khalifah_et_al_2004} 
all of which display a power-law $T$-dependence of $\delta\rho$ with an exponent $s$ that is clearly smaller than 2 on either side of the
quantum FM transition. Another example is the HM MnSi, which is tunable by pressure and displays a $T^{3/2}$ 
behavior over a temperature range of nearly three decacdes in a large region in the disordered phase, with a remarkably large
prefactor $A_{3/2}$, but $T^2$ behavior in the ordered phase.\cite{Pfleiderer_Julian_Lonzarich_2001} 
These systems all fulfill two requirements that help rule out possible reasons for the NFL behavior: (1) They are relatively clean, 
as evidenced by a small value of $\rho_0$, or a large mean-free path. This rules out effects due to diffusive electron dynamics that can lead to NFL 
behavior.\cite{Lee_Ramakrishnan_1985, Belitz_Kirkpatrick_1994} (2) The NFL behavior is generic, i.e., it is observed in large regions 
of the phase diagram, which rules out the possible influence of quantum critical behavior. In Figs.~\ref{fig:1}, \ref{fig:2} we show the
schematic phase diagrams of ZrZn$_2$ and MnSi, respectively, for illustrative purposes. Notice that the exponent $s$ for ZrZn$_2$ takes on its
smallest value of 1.5 in a pocket around $T = 10$K and $p = 2$GPa, but is close to 1.7 in a large part of the phase diagram. This somewhat ill-defined
value of $s$ suggests that several competing scattering mechanisms lead to an effective power law. In MnSi, by contrast, a very clean
$T^{3/2}$ behavior is observed over more than two temperature decades, which suggests one dominating scattering mechanism.
There are other examples, such as Nb$_{1-y}$Fe$_{2+y}$,\cite{Brando_et_al_2008} NiGa$_3$, \cite{Fluitman_et_al_1973, Pfleiderer_2007}
and (Ni$_{1-x}$Pd$_x$)$_3$Al,\cite{Sato_1975} that allow for more possibilities since they lack one or 
both of these requirements; we will come back to some of these in the discussion. 

In the context of requirement (2), and for later reference, we mention that it is now well understood 
theoretically,\cite{Belitz_Kirkpatrick_Vojta_1999,Kirkpatrick_Belitz_2012b} and confirmed 
experimentally,\cite{Brando_et_al_2016a} that the quantum phase transition in clean metallic FMs is generically 
discontinuous or first order, so by definition there is no standard magnetic quantum critical behavior. This is indeed the 
case for all of the FM examples listed above, and also for MnSi. Requirement (2) is
still useful, however, as critical fluctuations may be found in a pre-asymptotic region in the vicinity of a transition that is
weakly first order. We also mention that phase separation is routinely observed away from the coexistence curve in
systems that display a first-order transition,\cite{Uemura_et_al_2007, Pfleiderer_et_al_2010, Gat-Malureanu_et_al_2011}
which means that each phase displays, to some extent, characteristics of the other phase. 
We will come back to this phenomenon, as it is crucial for some of our arguments. 

The generic NFL behavior summarized above has proved very challenging to explain, and is still far from understood. One mechanism leading to a
 $T^{3/2}$ behavior that is applicable to HMs was reported in Ref.~\onlinecite{Kirkpatrick_Belitz_2010}, where it was shown that scattering of
electrons by columnar fluctuations (which in HMs can be realized by skyrmionic spin textures, see Ref.~\onlinecite{Muehlbauer_et_al_2009}) in conjunction with 
weak quenched disorder leads to a $T^{3/2}$ behavior of the electrical resistivity.

The fact that similar NFL behavior has been found in various FMs implies that the above mechanism, which
vanishes in the FM limit $q\to 0$, is not sufficiently general to explain all of the observations. In addition, the
exponent observed in MnSi is $s=1.5$ over several decades of temperatures,\cite{Pfleiderer_Julian_Lonzarich_2001}
whereas the exponents observed in FMs, while clearly less than $2$, tend to be slightly larger than 1.5, and the
smallest values of $s$ are not necessarily observed at the lowest temperatures, see Fig.~\ref{fig:1}. Also, 
in MnSi the NFL behavior is observed in the disordered phase only, whereas in the various ferromagnets it is found
in either phase. All of this indicates that different mechanisms may be at work in different materials. The observed NFL transport
properties of low-temperature FMs and HMs thus remains a major unsolved problem. 

In this paper we derive and discuss three additional mechanism that lead to a $T^{3/2}$ NFL behavior of the electrical resistivity.
The first one relies on FM order and involves magnon-mediated scattering of electrons in different subbands of the exchange-split
conduction band (interband scattering) in the presence of weak disorder. The second mechanism is applicable to
HMs with skyrmionic spin textures, where interband scattering in conjuction with weak disorder leads, in a large pre-asymptotic
temperature region, to a $T^{3/2}$ behavior. The third one involves HMs without skyrmions and 
does not require any quenched disorder. It involves scattering of electrons in different subbands
with the interaction mediated by helimagnons. We compare and contrast these mechanisms with the one
previously reported in Ref.~\onlinecite{Kirkpatrick_Belitz_2010}, and also with various competing scattering
mechanisms that lead to different power laws.

This paper is organized as follows: In the remainder of this section we summarize our main results. In Sec.~\ref{sec:II} we discuss
some general physical principles that allow for very simple derivations of our results, which are given in Sec.~\ref{sec:III}. We discuss our results
and the experimental situation in Sec.~\ref{sec:IV}.

\subsection{Summary of main results}
\label{subsec:I.B}

For the convenience of the reader we first summarize our main results; see also Table~\ref{table:1}. We have identified three mechanisms that lead to a $T^{3/2}$ 
behavior of the electrical resistivity in sizable temperature regimes, namely:

\subsubsection{Mechanism 1: Magnon-mediated scattering in weakly disordered ferromagnets}
\label{subsubsec:I.B.1}

The first mechanism relies on FM order. It involves magnon-mediated scattering of electrons in different subbands of the exchange-split conduction band
(interband scattering) in the presence of weak disorder (ballistic regime, see Refs.~\onlinecite{Zala_Narozhny_Aleiner_2001} and \onlinecite{Kirkpatrick_Belitz_Saha_2008a,
Kirkpatrick_Belitz_Saha_2008b}). 
In the current context, weak disorder is defined by the inequalities $\lambda\tau \gg 1$ and  $T \gg T_{\text{ball}} \equiv T_1/(\epsilonF\tau)^2$,
with $\lambda$ the exchange splitting of the conduction band, $T_1$ the magnon energy at the edge of the Brillouin zone (i.e., the magnetic Debye
temperature), $\epsilonF$ the Fermi energy, and $\tau$ the elastic mean-free time that determines the residual
resistivity. The result for $\delta\rho$ for this case can be written
\be
\delta\rho_{\text{FM}}/\rho_0 = \gamma_1 T^{3/2}/T_1\sqrt{T_0}\ ,
\label{eq:1.2}
\ee
with $\gamma_1$ a numerical constant. It is valid for $T_{\text{ball}} \ll T \alt T_0$, with $T_0 = T_1 \lambda^2/\epsilonF^2$.
Equation~(\ref{eq:1.2}) also holds for the thermal resistivity, only the numerical prefactor is different.

\subsubsection{Mechanism 2: Skyrmionic columnar fluctuations in helimagnets with weak disorder}
\label{subsebsec:I.B.2}

The second mechanism is applicable to HMs with skyrmionic spin textures comprised of three helices with pitch wave number $q$, as proposed in
Ref.~\onlinecite{Muehlbauer_et_al_2009}. Fluctuations of the resulting columnar order lead, in conjuction with weak
disorder and for interband scattering, to a result that is identical to the one in weakly disordered FMs,
\be
\delta\rho_{\text{sky}}/\rho_0 = \gamma_2 T^{3/2}/T_1\sqrt{T_0}\ ,
\label{eq:1.3}
\ee
with $\gamma_2$ another constant. The prefactor of the $T^{3/2}$ law is much larger than the one found
previously for intraband scattering mediated by columnar fluctuations.\cite{Kirkpatrick_Belitz_2010} This result is valid 
for $\Tball \ll T \alt T_0$, except that for temperatures lower than $T_1(q/\kF)^4$ it crosses over to a $T^{5/4}$ behavior (provided 
the crossover temperature is larger than $\Tball$). It also holds for both the electrical and 
the thermal resistivity. 

\subsubsection{Mechanism 3: Helimagnon-mediated scattering in clean helimagnets}
\label{subsubsec:I.B.3}

The third mechanism is applicable to HMs, and does not rely on quenched disorder. It
involves interband scattering of electrons mediated by helimagnons.
While the dominant contribution at asymptotically low $T$ is given by the scattering of electrons in the
same subband (intraband scattering, see Refs.~\onlinecite{Belitz_Kirkpatrick_Rosch_2006b, Kirkpatrick_Belitz_Saha_2008b}),
this mechanism provides the leading contribution in a pre-asymptotic temperature window. The result is
\be
\delta\rho_{\text{HM}}/\rho_0 = \gamma_3 \left(q/\kF\right) \lambda\tau \left(T/T_1\right)^{3/2}\ .
\label{eq:1.4}
\ee
Here $q$ is the pitch wave number of the helically ordered phase, 
and $\gamma_3$ is a third constant. Equation~(\ref{eq:1.4}) is valid for $q/\kF \ll 1$ and $T_0 \alt T \alt T_1 (q/\kF)^2$
(provided this temperature window exists).  
This result holds for the electrical resistivity, the contribution to the thermal resistivity is proportional to $T$.

\section{General considerations}
\label{sec:II}

All of the mechanisms discussed in this paper 
hinge on three basic observations. The first one is that any nonanalytic $T$-dependence of $\delta\rho$ requires the existence of
soft or massless modes that are infinitely long lived in the limit of long wavelengths and couple to the conduction electrons. 

The second observation is that in any metallic magnet the Fermi surface is split by the exchange splitting $\lambda$. In general one thus expects two types of
scattering processes. One is scattering between electrons in different subbands (interband scattering). In clean systems, these processes will be exponentially frozen
out at low temperatures, so any power law generated by an exchange of soft modes must cross over to an exponentially small rate at asymptotically
low temperatures. A second type of processes is scattering between electrons in the same subband (interband scattering). These can produce power
laws even at arbitrarily low temperatures, and hence will asymptotically always dominate the interband processes. However, as we will see, this
argument may be irrelevant for practical purposes, since small prefactors can make the asymptotic low-temperature region unobservably small,
and small amounts of quenched disorder can qualitatively change the behavior.
Also, in ferromagnets intraband scattering is absent since ferromagnetic magnons do not couple electrons in the same subband.

The third observation is that in the vicinity of a first-order transition, as is realized at low $T$ in all
of the magnets under discussion (see above), phase separation, i.e., the existence of a finite fraction of the
ordered phase within the disordered one, and vice versa, is commonly observed; see Refs.~\onlinecite{Uemura_et_al_2007,
Pfleiderer_et_al_2010, Gat-Malureanu_et_al_2011} for examples involving some of the magnets under discussion here. This has recently been
explained by the realization that static, finite-size droplets of the minority phase are stable within the
majority phase on either side of the coexistence curve, provided there is a moderate amount of quenched
disorder that couples predominantly to the order-parameter degrees of freedom, in this case the 
magnetization.\cite{Kirkpatrick_Belitz_2016b} This allows features that are characteristic of one phase
to exist, to a limited extent, on both sides of the coexistence curve. This will play an important role in
our discussions.

We now discuss the relevant soft modes and energy scales, and provide basic expressions for scattering rates that can be used
to derive all of our results.

\subsection{Goldstone modes, and susceptibilities}
\label{subsec:II.A}

An 
obvious soft mode in a FM is the ferromagnon with dispersion relation \cite{Kittel_2005}
\begin{equation}
\omega_{\text{FM}}(\bm{k}\rightarrow 0) = D {\bm k}^2
\label{eq:2.1}
\end{equation}
with $D$ the spin-stiffness coefficient.

The corresponding Goldstone mode in the helical phase of a HM is the helimagnon with dispersion relation \cite{Belitz_Kirkpatrick_Rosch_2006a}
\bse
\label{eqs:2.2}
\be
\omega_{\text{HM}}({\bm k}\to 0) = \sqrt{c_2 k_z^2 + c_4{\bm k}_{\perp}^{4}}\ ,
\label{eq:2.2}
\ee
with $k_z$ and ${\bm k}_{\perp}$ the components of ${\bm k}$ parallel and perpendicular, respectively, to the pitch wave
vector ${\bm q}$. In the limit of a long pitch wavelength, $q/k_{F}\ll1$, one has $c_{2}\propto c_{4}\,q^{2}$ with $c_{4}\propto D^{2}$. 
Ignoring numerical prefactors, we have
\be
\omega_{\text{HM}}({\bm k}) = D \sqrt{q^2 k_z^2 + {\bm k}_{\perp}^4}\quad (k \alt q) \ .
\label{eq:2.2b}
\ee
\ese
This is valid for $k \alt q$; in the opposite limit the resonance frequency crosses over to the ferromagnetic one.

Finally, in a phase with columnar order the Goldstone modes are columnar fluctuations with a dispersion relation of the form\cite{DeGennes_Prost_1993, Ho_et_al_2010}
\bse
\label{eqs:2.3}
\be
\omega_{\text{col}}({\bm k}) = \sqrt{c_4 k_z^4 + c_{2}{\bm k}_{\perp}^2 }\ .
\label{eq:2.3a}
\ee
An example of columnar order in a magnet is the skyrmion spin texture observed in the A-phase of the HM 
MnSi.\cite{Muehlbauer_et_al_2009}  If the skyrmions are comprised by a superposition of three helices, as proposed in Ref.~\onlinecite{Muehlbauer_et_al_2009}, one has
again $c_2 \propto c_4 q^2 \propto D^2 q^2$, and ignoring numerical prefactors we write
\be
\omega_{\text{sky}}({\bm k}) = D \sqrt{k_z^4 + {\bm k}_{\perp}^2 q^2}\quad (D q^4/\kF^2 \alt \omega_{\text{sky}} \alt D q^2)\ .
\label{eq:2.3b}
\ee
The region of validity of this expression is bounded above by $k \approx q$ or $\omega \alt Dq^2$; for larger $k$ the frequency again crosses
over to the ferromagnetic one. It is bounded below by $\omega \agt D q^4/\kF^2$; for asymptotically small frequencies
the dynamics of the soft modes change and one obtains\cite{Petrova_Tchernyshyov_2011}
\be
\omega_{\text{sky}}({\bm k}) = D(\kF^2/q^2) \left(k_z^4/q^2 + {\bm k}_{\perp}^2\right) \quad (\omega_{\text{sky}} \alt D q^4/\kF^2)\ .
\label{eq:2.3c}
\ee
\ese

\begin{figure}[t]
\includegraphics[width=6cm]{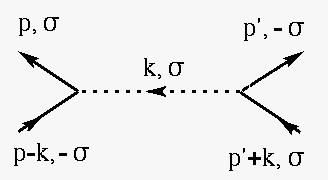}
\caption{Effective interband scattering of electrons mediated by magnon exchange. The effective potential
              is represented by the dotted line, and $\sigma = \pm 1 \equiv \uparrow,\downarrow$ is the spin projection.}
\label{fig:3}
\end{figure}

We will refer to all of these Goldstone modes arising from magnetic order summarily as ``magnons''. For deriving their contributions
to the electronic relaxation rates, we consider the effective interaction between electrons mediated by the exchange
of magnons, see Fig.~\ref{fig:3}. 
The effective potential is proportional to the susceptibility $\chi$ of the order-parameter phase fluctuations that also determine
the Goldstone propagator. In the case of the FM these phase fluctuations are simply the transverse components of the magnetization. In
the HM case they are a generalized phase that involves a transverse gradient that leads to the characteristic anisotropic
dispersion relation for the helimagnons.\cite{Belitz_Kirkpatrick_Rosch_2006a} 
The leading contribution to $\chi$ in FMs and in the helical and columnar phases of HMs,
respectively, apart from a numerical prefactor, is given by the following expressions 
(see Refs.~\onlinecite{Bharadwaj_Belitz_Kirkpatrick_2014, Belitz_Kirkpatrick_Rosch_2006b}, and also Sec.~\ref{subsec:IV.A}).
\smallskip
\bse
\label{eqs:2.4}

\noindent FM:
\be
\quad\chi({\bm k},i\Omega) \propto \frac{m_0 D{\bm k}^2}{\omega_{\text{FM}}({\bm k})^2 - (i\Omega)^2}\ ,
\label{eq:2.4a}
\ee

\noindent
HM helical:
\be
\chi({\bm k},i\Omega) \propto \frac{m_0 D{\bm q}^2}{\omega_{\text{HM}}({\bm k})^2 - (i\Omega)^2}\ ,
\label{eq:2.4b}
\ee

\noindent
HM skyrmionic:
\be
\chi({\bm k},i\Omega) \propto \begin{cases} \frac{m_0 D{\bm q}^2}{\omega_{\text{sky}}({\bm k})^2 - (i\Omega)^2}\hskip 20pt \text{for}\ \omega_{\text{sky}}\agt D q^4/\kF^2 \\
                                                                          \frac{m_0 (\kF/q)^2 \omega_{\text{sky}}({\bm k})}{\omega_{\text{sky}}({\bm k})^2 - (i\Omega)^2} \quad \text{for}\ 
                                                                                                                                                                                                           \omega_{\text{sky}}\alt D q^4/\kF^2\ .
                                                    \end{cases}
\label{eq:2.4c}
\ee
\ese
Here $m_0$ is the magnetization scale, i.e., the expectation value of the local spin density, and we neglect the damping of the magnons. 
Note that in a FM the numerator is proportional to a gradient squared, while in the HM helical case this is replaced by a $q^2$. As a result, 
the HM helical susceptibility is softer than the FM one, even though the HM helical resonance frequency is stiffer than the FM one.

\subsection{Energy scales}
\label{subsec:II.B}

For later reference we discuss various energy scales that are relevant for the problem at hand. The highest of these is the microscopic
or atomic energy scale $E_a$. In good metals this is usually the Fermi energy,\cite{microscopic_scales_footnote} which we denote by 
$\epsilonF$. The magnetism is characterized by
several closely related energy scales. One is given by the maximum excitation energy, 
\be
T_1 = D/a^2 \approx D\kF^2\ , 
\label{eq:2.5}
\ee
with $a$ the microscopic length scale, which in good metals is close to the inverse of the Fermi wave number $\kF$. $T_1$ is the magnetic analog
of a Debye temperature. A second one is the exchange splitting $\lambda$, which is due to the effective magnetic field seen by the
conduction electrons. It is related to the magnetization scale $m_0$ via the exchange interaction $\Gamma_t$ that couples the electronic
spin to the magnetization:\cite{energy_scales_footnote}
\be
\lambda = \Gamma_t m_0 \ .
\label{eq:2.6}
\ee

A third relevant scale arises in interband scattering processes. The smallest wave number that can be transferred by means
of magnon exchange is on the order of $k_0 = \lambda/\vF$, with $\vF$ the Fermi wave number, and the smallest energy that can
be transferred is therefore
\be
T_0 = D k_0^2 \approx T_1\left(\lambda/\epsilon_F\right)^2\ .
\label{eq:2.7}
\ee

Another energy scale is relevant in HMs. The expression (\ref{eq:2.2b}) is valid only for $k<q$; for larger
wave numbers they cross over to a ferromagnetic dispersion relation. The temperature region where these excitations
determine the scaling of the scattering rates is thus bounded above by 
\be
T_q = D q^2 = T_1 (q/\kF)^2\ .
\label{eq:2.8}
\ee

Finally, consider quenched disorder as characterized by an elastic scattering time $\tau$. We will be interested in the
ballistic regime,\cite{Zala_Narozhny_Aleiner_2001} which is separated from the diffusive regime by the requirement
$\vF k > 1/\tau$, with $k$ the momentum transfer in the scattering process (or, alternatively, by $k\ell > 1$, with $\ell$
the elastic mean-free path). The magnon frequency scales as the
temperature, and from Eqs.~(\ref{eq:2.1}, \ref{eq:2.2b}, \ref{eq:2.3b}) we see that in all of these cases the wave number squared scales
as the magnon frequency divided by $D$, i.e., the wave number scales as $k \sim \sqrt{T/D}$. The ballistic transport regime
is therefore restricted to temperatures $T > T_{\text{ball}}$, where
\be
T_{\text{ball}} = T_1/(\epsilonF\tau)^2\ .
\label{eq:2.9}
\ee
We will be interested in situations where $T_0$ and $T_{\text{ball}}$ are well separated, $T_{\text{ball}} \ll T_0$, which
requires the weak-disorder condition $\lambda\tau \gg 1$.

\subsection{General expressions for the scattering rates}
\label{subsec:II.C}

We now provide expressions for the scattering rates in terms of integrals, first for the single-particle rate $1/\tau_{\text{sp}}$ in clean systems, for both interband and intraband
scattering. $1/\tau_{\text{sp}}$ determines the thermal resistivity.\cite{Wilson_1954} The electrical resistivity is determined by the
transport rate $1/\tau_{\text{tr}}$. It is well known that in clean systems the latter differs from the former by a factor in the integrand that suppresses backscattering
and is proportional to the momentum transfer squared, or $({\bm k}-{\bm p})^2/\kF^2$ in Eqs.~(\ref{eq:2.10}, \ref{eq:2.11}) below.\cite{Wilson_1954} We will 
later use this observation to deduce the transport rate from the single-particle one. Together with simple arguments about the effects of disorder this will suffice to 
obtain both rates, and hence both the thermal and the electrical resistivity, for all cases of interest by just considering the single-particle rate in clean systems. 

\subsubsection{Single-particle rate, clean systems}
\label{subsubsec:II.C.1}

\paragraph{a) Interband scattering}
\label{par:II.C.1.a}

\begin{figure}[t]
\includegraphics[width=6cm]{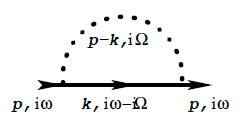}
\caption{Exchange contribution to the single-particle self energy that results from the interaction vertex shown in Fig.~\ref{fig:3}.}
\label{fig:4}
\end{figure}
We start with an expression for the single-particle relaxation rate $1/\tau_{\text{sp}}$ due to magnon exchange between electrons in different subbands of the
exchange-split conduction band (interband scattering) in clean systems. For a quasiparticle with wave vector ${\bm k}$ it is given by the imaginary part
of the Fock or exchange contribution to the quasiparticle self energy shown in Fig. \ref{fig:4}. Averaging over the Fermi surface, and
remembering that the effective potential is proportional to the phase-fluctuation susceptibility, we find, apart from purely numerical prefactors,
\bea
\frac{1}{\tau_{\text{sp}}} &\propto& \NF\Gamma_t^2 \int du\ \frac{1}{\sinh(u/T)} \frac{1}{\NF^2 V^2}\sum_{{\bm k},{\bm p}} \delta(\xi_{\bm k}-\lambda)\,
\nonumber\\
&&\hskip 60pt \times\delta(\xi_{\bm p}+\lambda)\, \chi''({\bm k}-{\bm p},u)
\label{eq:2.10}
\eea 
Here $\xi_{\bm k} = 0$ defines the Fermi surface in the absence of an exchange splitting,
$V$ is the system volume, and $\chi''$ is the spectrum of the susceptibility given in
Eqs.~(\ref{eqs:2.4}). The two $\delta$-functions pin the electrons to the respective Fermi surfaces. With $\chi$ representing the FM susceptibility,
Eq.~(\ref{eq:2.4a}), this expression was considered in Ref.~\onlinecite{Bharadwaj_Belitz_Kirkpatrick_2014}; it also holds for the two helical cases.

\paragraph{b) Intraband scattering}
\label{par:II.C.1.b}

We now consider magnon exchange between electrons in the same subband (intraband scattering). This is relevant only for the helical cases;
in ferromagnets the magnons do not couple electrons in the same subband. The expression for the single-particle relaxation rate was derived in
Refs.~\onlinecite{Belitz_Kirkpatrick_Rosch_2006a} and \onlinecite{Kirkpatrick_Belitz_Saha_2008a}. The result is very similar to Eq.~(\ref{eq:2.10});
it takes the form
\bea
\frac{1}{\tau_{\text{sp}}} &\propto&\NF \Gamma_t^2 \left(\frac{q}{\kF}\right)^2 \left(\frac{\epsilonF}{T_1}\right)^2 \int du\ \frac{1}{\sinh(u/T)} \frac{1}{\NF^2 V^2}\sum_{{\bm k},{\bm p}} \nonumber\\
&&\hskip 0pt \times\,\delta(\xi_{\bm k})\, \delta(\xi_{\bm p})\, \frac{({\bm k}-{\bm p})^2}{\kF^2}\, \chi''({\bm k}-{\bm p},u)
\label{eq:2.11}
\eea 
The $\delta$-functions now reflect the fact that both electrons belong to the same Fermi surface. The extra factor of $({\bm k}-{\bm p})^2$ compared
to Eq.~(\ref{eq:2.10}) is due to the fact that electrons in the same band can couple only to gradients of the fluctuating phase, whereas electrons in
different bands couple to the phase directly.\cite{gradient_coupling_footnote}

\subsubsection{Transport rate, clean systems}
\label{subsubsec:II.C.2}

A determination of the electrical resistivity requires solving the Boltzmann equation, or, equivalently, evaluating the Kubo formula. Even in the
simplest conserving approximation this requires solving an integral equation for a vertex function. This integral equation is usually replaced by
an algebraic equation.\cite{Mahan_1981} In Ref.~\onlinecite{Belitz_Kirkpatrick_2010a} it was shown that this approximation is exact with respect
to the leading low-temperature dependence of the resistivity, and the resulting algebraic equations for FMs were derived and solved in
Ref.~\onlinecite{Bharadwaj_Belitz_Kirkpatrick_2014}. The final result is as follows: The electrical resistivity is effectively given by the transport rate $1/\tau_{\text{tr}}$, 
which is obtained from the same integral as the single-particle rate, Eqs.~(\ref{eq:2.10}, \ref{eq:2.11}), but with an additional factor of 
$({\bm k} - {\bm p})^2/\kF^2$ in the integrand. In the context of the Boltzmann equation this is known as the backscattering factor; it
suppresses large-angle scattering and thus weakens the temperature dependence of the transport rate compared to the single-particle rate.\cite{Wilson_1954}
With the exception of the skyrmionic case at asymptotically low frequencies, the wave number squared scales as the Goldstone-mode frequency
for all of the cases we are considering, see Sec.~\ref{subsec:II.A}, which in turn scales
as the temperature. We can thus immediately anticipate that generically the leading $T$-dependence of the transport rate will have an additional
factor of $T$ compared to the single-particle rate. The only caveat is that this argument assumes that the final integrals are infrared convergent once
the temperature has been scaled out; this is not always the case, see Sec.~\ref{subsec:III.C} below.

\subsubsection{Effects of disorder}
\label{subsubsec:II.C.3}

\begin{figure}[t]
\includegraphics[width=4cm]{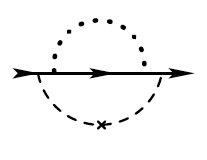}
\caption{Leading contributions to the electron self energy in the weak-disorder or ballistic regime. The dotted line represents the effective
              interaction mediated by the magnons; the dashed line with a cross represents the quenched impurities. }
\label{fig:5}
\end{figure}

\begin{figure}[t]
\includegraphics[width=8cm]{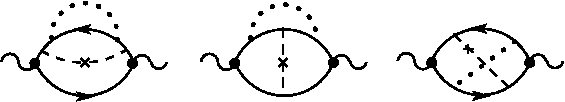}
\caption{Leading contributions to the conductivity in the weak-disorder or ballistic regime. Dotted lines represent the effective
              interaction mediated by the magnons; dashed lines with crosses represent the quenched impurities. The dots represent
              the external current vertices.}
\label{fig:6}
\end{figure}

The effects of weak quenched disorder are qualitatively different depending on whether we consider interband or intraband scattering. A
complete discussion requires evaluating the Kubo formula with impurity scattering taken into account. The leading contribution to the
electron self energy in the weak or ballistic disorder regime, $\lambda\tau \gg 1$, is shown in Fig.~\ref{fig:5}, and the leading contributions to the
electrical conductivity are shown in Fig.~\ref{fig:6}; see Refs.~\onlinecite{Kirkpatrick_Belitz_Saha_2008a, Kirkpatrick_Belitz_Saha_2008b} for a 
complete discussion for the case of helimagnon-mediated scattering. However, for the cases of
interest here a much simpler argument suffices; we discuss its relation to the detailed calculation in the Appendix~\ref{app:A}.

\paragraph{a) Interband scattering}
\label{par:II.C.3.a}

Consider the expression for the clean single-particle rate given in Eq.~(\ref{eq:2.10}). Shifting the momentum ${\bm k}$ by ${\bm p}$ and
performing the ${\bm p}$-integration yields
\bea
\frac{1}{\NF V}\sum_{\bm p} \delta(\xi_{{\bm k}+{\bm p}} - \lambda)\,\delta(\xi_{\bm p} + \lambda) &\propto& \int_{-1}^{1} d\eta\,\delta(k\vF\eta - 2\lambda)
\nonumber\\
&&\hskip -40pt = \frac{1}{\vF k}\,\Theta(k - 2\lambda/\vF)\ .
\label{eq:2.12}
\eea
Note the theta function, which leads to an exponentially small scattering rate at asymptotically low temperatures.\cite{Bharadwaj_Belitz_Kirkpatrick_2014}
Weak disorder smears out the $\delta$-function and we have, in the limit $\vF k \ll \lambda$ and $\lambda\tau \gg 1$,
\bea
\frac{1}{\vF k}\,\Theta(k - 2\lambda/\vF) &=& \int_{-1}^1 d\eta\,\delta(\vF k\eta - 2\lambda) 
\nonumber\\
&&  \hskip -80pt \rightarrow \int_{-1}^1 d\eta\ \frac{1/\tau}{(\vF k\eta - 2\lambda)^2 + 1/\tau^2} \approx \frac{1}{\lambda^2\tau}\ .
\label{eq:2.13}
\eea
The disorder thus results in, (1) an extra factor of $\vF k/\lambda^2\tau$ in the integrand, and (2) the elimination of the step function, and
hence of the lower cutoff for the $k$-integral. Since $k$ scales as $k \sim T^{1/2}$, this means that disorder leads to a power-law temperature
dependence of $1/\tausp$ that is {\em weaker} than in the clean case by a factor of $T^{1/2}$, but extends to temperatures below the energy
scale $T_0$. 

For the transport rate, disorder eliminates the backscattering factor since it leads to more isotropic scattering. The effective extra factor in the
integrand is thus $(\epsilonF/\lambda^2\tau)\kF/k$, and disorder {\em strengthens} the temperature dependence of the rate by a factor of $1/T^{1/2}$.
As a result, the single-particle rate and the transport rate display the same temperature dependence. Again, these conclusions may be
modified by the convergence properties of the final dimensionless integrals.

\paragraph{b) Intraband scattering}
\label{par:II.C.3.b}

In the case of intraband scattering the arguments of the $\delta$-functions in Eq.~(\ref{eq:2.12}) are not shifted with respect to one another,
and the smearing argument yields
\bea
\frac{1}{\vF k} &=& \int_{-1}^1 d\eta\,\delta(\vF k\eta) 
\nonumber\\
&&  \hskip -80pt \rightarrow \int_{-1}^1 d\eta\ \frac{1/\tau}{(\vF k\eta)^2 + 1/\tau^2} \approx \frac{1}{\vF k}\left[1 - \frac{1}{\vF k\tau}\right]\ .
\label{eq:2.14}
\eea
Disorder thus provides a correction to the scattering rate that comes with an extra factor of $1/\vF k\tau$ in the integrand and {\em strengthens}
the temperature dependence of $1/\tausp$ by a factor of $1/T^{1/2}$. The sign of this contribution is negative, and it always is a small correction
to the clean contribution (since $T>\Tball$). For the transport rate the backscattering factor is suppressed in addition, and disorder strengthens
the $T$-dependence of the rate by a factor of $1/T^{3/2}$. The disorder corrections to both rates thus again have the same temperature
dependence. 

The above simple argument always yields the correct temperature scaling of both rates for all of the cases considered here. However, the
sign of the disorder correction is rendered correctly only for the skyrmionic case, whereas intraband scattering by helimagnons results in
a positive disorder correction to the rates.\cite{Kirkpatrick_Belitz_Saha_2008a, Kirkpatrick_Belitz_Saha_2008b} The reason is that in this
case the simple factor of $({\bm k} - {\bm p})^2$ in Eq.~({\ref{eq:2.11}) does not adequately describe the coupling of the Goldstone modes 
to the electrons. The actual coupling contains an angular dependence that flips the sign, as explained in the Appendix~\ref{app:A}.

All of these observations are consistent with the results of the explicit calculations in Refs.~\onlinecite{Kirkpatrick_Belitz_Saha_2008a,
Kirkpatrick_Belitz_Saha_2008b}.

\section{Derivations}
\label{sec:III}

We are now in a position to provide very simple derivations of the contributions to the scattering rates, and hence the electrical and
thermal resistivities, due to the exchange of ferromagnons, skyrmionic columnar fluctuations, or helimagnons, with or without
weak quenched disorder, and involving either interband or intraband scattering.

\subsection{Mechanism 1: Magnon-mediated scattering in weakly disordered ferromagnets}
\label{subsec:III.A}

For ferromagnets only the interband scattering mechanism is relevant; the magnons do not couple electrons in the same
subband. For clean systems the single-particle rate is given by Eq.~(\ref{eq:2.10}). With Eqs.~(\ref{eq:2.4a}) and (\ref{eq:2.1})
for the susceptibility and the magnon frequency, respectively, we have
\bea
\frac{1}{\tau_{\text{sp}}} &\propto& \frac{\lambda}{\NF V}\sum_{\bm k} \frac{1}{\sinh(D k^2/T)} \int_{-1}^1 d\eta\ \delta(k\vF\eta - 2\lambda)
\nonumber\\
&=&\frac{\lambda}{\NF V} \sum_{\bm k} \frac{1}{\sinh(D k^2/T)}\,\frac{1}{\vF k}\,\Theta(k - 2\lambda/\vF)
\nonumber\\
&\propto& \frac{T\lambda}{T_1}\int_{T_0/T}^{T_1/T} dx\,\frac{1}{\sinh x}\ ,
\label{eq:3.1}
\eea
where we have dropped a factor of $\NF\Gamma_t = O(1)$, as we have all other numerical prefactors. We thus reproduce the
result of Ref.~\onlinecite{Bharadwaj_Belitz_Kirkpatrick_2014}:
\be
\frac{1}{\tau_{\text{sp}}} \propto (T\lambda/T_1)\times\begin{cases} e^{-T_0/T} \quad\ \ \, \text{for} \quad T \alt T_0 \\ 
                                                                               \ln(T/T_0) \quad \text{for}\quad T_0 \alt T \alt T_1\ .
                                                       \end{cases}
\label{eq:3.2}                                                       
\ee   
The corresponding result for $1/\tau_{\text{tr}}$ is $T^2\lambda/T_1^2$ for $T_0 \alt T \alt T_1$, and an exponentially small 
expression for $T \alt T_0$.\cite{Bharadwaj_Belitz_Kirkpatrick_2014}                                                    
Now consider weak disorder, characterized by $\lambda\tau \gg 1$. As explained in Sec.~\ref{subsubsec:II.C.3}, the integrand
in Eq.~(\ref{eq:3.1}) acquires, in the region $\vF k \ll \lambda$, an additional factor of $\vF k/\lambda^2\tau$, and the step
function disappears. Ballistic disorder thus leads to an additional contribution to the single-particle rate
\bse
\label{eqs:3.3}
\bea
\delta(1/\tau_{\text{sp}}) &\propto& \frac{1}{\lambda\tau \NF}\,\frac{1}{V}\sum_{\bm k} \frac{1}{\sinh(D k^2/T)}\,\Theta(\lambda - k\vF)
\nonumber\\
&\propto& \frac{1}{\tau}\,\frac{\epsilonF}{\lambda}\left(\frac{T}{T_1}\right)^{3/2} \int_0^{T_0/T} dx\,\frac{\sqrt{x}}{\sinh x}\ .
\label{eq:3.3a}
\eea
For $T\gg T_0$ this gives a small correction to the clean rate, Eq.~(\ref{eq:3.2}), but for $T\alt T_0$ it provides the leading contribution,
which is proportional to $T^{3/2}$. The result for the transport rate is the same, as explained in Sec.~\ref{subsubsec:II.C.3}, and
we obtain 
\be
\delta(1/\tau_{\text{tr}}) \propto \delta(1/\tau_{\text{sp}}) \propto \frac{1}{\tau}\,\frac{\epsilonF}{\lambda}\left(\frac{T}{T_1}\right)^{3/2}\quad (T \alt T_0)\ .
\label{eq:3.3b}
\ee
\ese
This is equivalent to Eq.~(\ref{eq:1.2}), which holds for both the electrical resistivity, which is given by $1/\tau_{\text{tr}}$, and the
thermal resistivity, which is given by $1/\tau_{\text{sp}}$.

These results are summarized in Table~\ref{table:1}.

\subsection{Mechanism 2: Skyrmionic columnar fluctuations in helimagnets with weak disorder}
\label{subsec:III.B}

Columnar fluctuations of any kind have a resonance frequency of the form given in Eq.~(\ref{eq:2.3a}). If the columns are formed
by skyrmionic spin textures that result from a superposition of three helices with pitch wave number $q$, as was proposed in
Ref.~\onlinecite{Muehlbauer_et_al_2009}, then the resonance frequency is given by Eq.~(\ref{eq:2.3b}) in the sizable frequency
window $T_q\, q^2/\kF^2 \alt \omega \alt T_q$, and by Eq.~(\ref{eq:2.3c}) for asymptotically small frequencies or wave numbers. 
The relevant susceptibility in these two regimes is given by Eq.~(\ref{eq:2.4c}).
Repeating the calculation from Sec.~\ref{subsec:III.A} we then obtain for the single-particle rate due to interband
scattering in clean systems
\bea
\frac{1}{\tau_{\text{sp}}} &\propto& \frac{T\lambda}{T_1} \int_{\sqrt{T_0/T}}^{\sqrt{T_1/T}} \frac{dz}{z} \int_0^{\infty} dx\,\frac{1}{\sqrt{z^4 + x}}\,\frac{1}{\sinh\sqrt{z^4 + x}}
\nonumber\\
&\propto& \begin{cases}  (T^2\lambda/T_0 T_1) e^{-T_0/T} \quad \text{for} \quad T\alt T_0 \\
                                        (T\lambda/T_1) \ln^2(T/T_0)\hskip 10pt \text{for}\quad T_0 \alt T \alt T_q \ .
                 \end{cases}
\label{eq:3.4}
\eea
The second temperature window may or may not exist, depending on the relative values of $T_0$ and $T_q$. For $T\agt T_q$ the
result crosses over to the FM one. The corresponding result for $1/\tau_{\text{tr}}$ is exponentially small and $T^2\lambda/T_1^2$, 
respectively, in the two temperature regions. In the presence of ballistic disorder both rates go as $T^{3/2}$ by the same arguments 
as in the FM case, and we obtain
\bea
\delta\left(1/\tau_{\text{tr}}\right) \propto \delta\left(1/\tau_{\text{sp}}\right) &\propto& \frac{1}{\tau}\,\frac{\epsilonF}{\lambda}
       \left(\frac{T}{T_1}\right)^{3/2}\quad 
\nonumber\\
&& \hskip -50pt  \left(\text{Max}(\Tball, T_q\, q^2/\kF^2\right) \alt T \alt T_q)\ .
\label{eq:3.5}
\eea
This result is equivalent to Eq.~(\ref{eq:1.3}). For $T \agt T_q$ it crosses over to the FM result, so ignoring the numerical
prefactor it effectively is valid for temperatures $T \alt \text{Max}(T_q, T_0)$. 

For asymptotically small frequencies or wave numbers the resonance frequency and the phase susceptibility are given by
Eq.~(\ref{eq:2.3c}) and the second expression in (\ref{eq:2.4c}), respectively. Repeating the calculation for this case we find
\bea
\delta\left(1/\tau_{\text{tr}}\right) \propto \delta\left(1/\tau_{\text{sp}}\right) &\propto& \frac{1}{\tau}\,\frac{\epsilonF}{\lambda}\,\frac{q}{\kF}\left(\frac{T}{T_1}\right)^{5/4} \quad 
\nonumber\\
&&\hskip -20pt ({\tilde T}_{\text{ball}} \alt T \alt T_q\, q^2/\kF^2)\ .
\label{eq:3.6}
\eea
Here ${\tilde T}_{\text{ball}} = \Tball\,(\kF/q)^4/(\epsilonF\tau)^2$.
We note that in HMs such as MnSi, $q/\kF$ tends to be on the order of $10^{-2}$. The temperature $T_q\,q^2/\kF^2$ where this asymptotic behavior
sets in therefore tends to be extremely low, and not necessarily larger than either $\Tball$ or ${\tilde T}_{\text{ball}}$, so the temperature window where Eq.~(\ref{eq:3.6})
is valid may not exist. 

For intraband scattering in clean systems, analogous considerations using Eq.~(\ref{eq:2.11}) readily show that the single-particle and transport rates
scale as $T^{2}$ and $T^3$, respectively, in the pre-asymptotic region $T_q\, q^2/\kF^2 \alt T \alt T_q$, which crosses over to $T^{3/2}$ and $T^2$,
respectively in the asymptotic region $\Tball \alt T \alt T_q\, q^2/\kF^2$. The disorder correction to either rate scales as $T^{3/2}$ in the pre-asymptotic
region, but it always is a small correction to the clean contribution, as explained in Sec.~\ref{subsubsec:II.C.3}. Asymptotically this crosses over to
a $T^{5/4}$ correction. 

All of these results are summarize in Table~\ref{table:1}.

\subsection{Mechanism 3: Helimagnon-mediated scattering in clean helimagnets}
\label{subsec:III.C}

In clean HMs, the leading helimagnon contribution to the relaxation rates at asymptotically low temperatures comes from interband scattering.
Using Eqs.~(\ref{eq:2.2b}) and (\ref{eq:2.4b}) in Eq.~(\ref{eq:2.11}) we reproduce the result of Ref.~\onlinecite{Belitz_Kirkpatrick_Rosch_2006b}
for the single-particle rate:
\be
\frac{1}{\tau_{\text{sp}}} \propto \lambda \left(\frac{q}{\kF}\right)^6 \left(\frac{\epsilonF}{T_1}\right)^2 \left(\frac{T}{T_q}\right)^{3/2}\ .
\label{eq:3.7}
\ee
The transport rate is given by the same expression with an additional factor of $T/T_1$. Ballistic disorder leads to corrections to both
rates that are proportional to $T$ and always small compared to the clean contribution.\cite{Kirkpatrick_Belitz_Saha_2008a, Kirkpatrick_Belitz_Saha_2008b} 
As explained in the Appendix~\ref{app:A}, the arguments given in Sec.~\ref{subsubsec:II.C.3} give the correct temperature scaling of the disorder corrections, 
but not the correct sign.

The interband contribution is exponentially small at asymptotically low temperatures, and therefore was not considered in Ref.~\onlinecite{Belitz_Kirkpatrick_Rosch_2006b}.
However, for values of the pitch wave number that are not too small it may dominate in a temperature window, as we now show. Using Eq.~(\ref{eq:2.11})
instead of (\ref{eq:2.10}) we obtain for the single-particle rate due to interband scattering
\bse
\label{eqs:3.8}
\bea
\frac{1}{\tau_{\text{sp}}} &\propto& \lambda\,\frac{q}{\kF}\,\left(\frac{T}{T_1}\right)^{1/2} \int_{T_0/T}^{T_1/T} \frac{dx}{\sqrt{x}} \int_0^{\infty} dz\, \frac{1}{\sqrt{z^2+x^2}}\,
\nonumber\\
&&\hskip 80pt \times  \frac{1}{\sinh{\sqrt{z^2 + x^2}}}\ .
\label{eq:3.8a}
\eea
The dimensionless integral provides an additional factor of $\sqrt{T/T_0}$, and we obtain
\be
\frac{1}{\tau_{\text{sp}}} \propto \lambda\,\frac{q}{\kF} \times \begin{cases} \sqrt{T/T_1}\,e^{-T_0/T}\quad \text{for}\quad T \alt T_0 \\
                                                                                                                       T/\sqrt{T_0 T_1} \hskip 33pt \text{for} \quad T_0 \alt T \alt T_q\ .
                                                                                               \end{cases}
\label{eq:3.8b}
\ee
\ese  
In the case of the transport rate the backscattering factor renders the dimensionless integral a constant, and we find
\be
\frac{1}{\tau_{\text{tr}}} \propto \lambda\left(\frac{q}{\kF}\right)^4 \left(\frac{T}{T_q}\right)^{3/2} \times \begin{cases} e^{-T_0/T}\quad \text{for} \quad T \alt T_0 \\
                                                                                                                                                                          1 \quad \text{for}\quad T_0 \alt T \alt T_q\ .
                                                                                                                                                    \end{cases}
\label{eq:3.9}
\ee   
For the electrical resistivity, this implies Eq.~(\ref{eq:1.4}). These results are valid for $T \alt T_q = D q^2$; for higher temperatures they cross over to the FM results. 
The temperature window where the resistivity scales as $T^{3/2}$ thus may or may not exist, depending on the value of the ratio $\epsilonF q/\lambda\kF$.     

As an alternative to the above derivation one can repeat the HM calculation of Ref.~\onlinecite{Belitz_Kirkpatrick_Rosch_2006b} or
\onlinecite{Kirkpatrick_Belitz_Saha_2008b} and take into account the interband scattering terms that were neglected in these
references. The result is the same. We also note that, while the resistivity in a HM is anisotropic, the
two independent components of the resistivity tensor differ only by a numerical factor.\cite{Kirkpatrick_Belitz_Saha_2008b}

In the presence of disorder the dimensionless integral is only logarithmically divergent in the infrared, rather than power-law divergent as in Eq.~(\ref{eq:3.8a}).
At the same time, the lower cutoff $T_0/T$ disappears and is replaced by $T_L/T$, where $T_L = D/L^2$ with $L$ the linear system size. We then obtain a
disorder correction to the rates, to leading logarithmic accuracy,
\bea
\delta(1/\tau_{\text{sp}}) \propto \delta(1/\tau_{\text{tr}}) &\propto& \frac{1}{\tau} \left(\frac{q}{\kF}\right)^3 \frac{\epsilonF}{\lambda}\,\frac{T}{T_q}\,\ln(T/T_L)
\nonumber\\
&& (\Tball \alt T \alt T_q)\ .
\label{eq:3.10}
\eea
This may or may not dominate over the clean contribution, depending on the parameter values. We note, however, that this result,
and in particular its remarkable logarithmic dependence on the system size, depend on the convergence properties of the dimensional
integral. A more detailed investigation of where the power-law divergence crosses over to a logarithmic one is therefore warranted in
order to determine the range of validity and the prefactor of the $T\ln T$ behavior.\cite{spin-orbit_footnote}
However, any such analysis will be highly model dependent, and for the purposes of the current general discussion we therefore defer to a future investigation.

\section{Summary, and Discussion}
\label{sec:IV}

Before we give a discussion of our results, we present a summary in the form of Table~\ref{table:1}. It lists the three process we have
identified that lead to a $T^{3/2}$ behavior of the electrical resistivity, as well as the other seven scattering processes we have discussed,
and also summarizes our results for the thermal resistivity.

We now discuss various aspects of our results. We first make some purely theoretical remarks, and then discuss issues relevant for
comparing with experiments. 

\begin{table*}[t]
\caption{Summary of temperature dependences of the single-particle scattering rate, or the thermal resistivity contribution $\delta\rho_{\text{th}}$, and the transport rate, or the
              electrical resistivity contribution $\delta\rho_{\text{el}}$, for different magnets and scattering mechanisms with or without ballistic disorder. Also shown is the prefactor
              $A_{3/2}$ for the electrical resistivity, $\delta\rho_{\text{el}}(T) = A_{3/2} T^{3/2}$, if applicable. $\rho_0 = m_e/n e^2\tau$ is the residual resistivity, and
              $\rho_{\lambda} = m_e\lambda/n e^2$. See the text for additional information.}
\vskip 10pt
\begin{ruledtabular}
\begin{tabular}{ c | c| Sc| Sc cccc }
System                  & Soft Modes                      & Scattering                       & Ballistic              & Single-particle                 & Transport                          & Prefactor & References  
\\[-5pt]
                              &                                         &                                        & Disorder             & Rate / $\delta\rho_{\text{th}}$   &  Rate / $\delta\rho_{\text{el}}$  & $A_{3/2}$&                     
\\
\cline{1-8}
\multirow{4}*{FM}  &                                         & \multirow{2}*{interband} & No$^{\,a)}$          & $T\ln (T_0/T)$                  & $T^2$                              &                     &        \onlinecite{Ueda_Moriya_1975, Bharadwaj_Belitz_Kirkpatrick_2014}                  
\\
                              &  magnons                       &                                       & Yes$^{\,b)}$         & $T^{3/2}$                          & $T^{3/2}$                        & $\rho_0 /T_1 T_0^{1/2}$ &  This work   
\\
\cline{3-8}
                              &                                        & intraband                       &                              &                                          &   N/A                               &                     &                
\\ 
\hline
\multirow{4}*{HM} & \multirow{5}*{skyrmions}  & \multirow{2}*{interband} & No$^{\,a),c)}$      &  $T\ln^2 (T/T_0) \to T$ & $T^{2} \to T^{3/2}$   &                  &   This work  
\\ 
                            &                                           &                                       & Yes$^{\,b),c)}$     &  $T^{3/2} \to T^{5/4}$   & $T^{3/2} \to T^{5/4}$ &  $\rho_0/T_1 T_0^{1/2}$&   This work  
\\
\cline{3-8}
                            &                                           & \multirow{2}*{intraband} & No$^{\,c)}$       & $T^2 \to T^{3/2}$        &  $T^3 \to T^{2}$                 &                  &   This work           
\\  
                            &                                           &                                        & Yes$^{\,d)}$     &  $(T^{3/2} \to T^{5/4})$   & $(T^{3/2} \to T^{5/4})$ &                   &   This work, \onlinecite{Kirkpatrick_Belitz_2010} 
\\
\hline
\multirow{5}*{HM}  &                                         & \multirow{2}*{interband} & No$^{\,a)}$         &  $T^{\,e)}$                   & $T^{3/2}$    &  $\rho_{\lambda}(q/\kF)/T_1^{3/2}$ &  This work   
\\
                             &  heli-                                 &                                        & Yes$^{\,b)}$        &  $T\ln(T/T_L)$               & $T\ln(T/T_L)$                   &                   &    This work  
\\
\cline{3-8}
                              &  magnons                         & \multirow{2}*{intraband$^{\,f)}$}  & No                      &  $T^{3/2}$                      &  $T^{5/2}$          &                   & \onlinecite{Belitz_Kirkpatrick_Rosch_2006b}                         
\\                           
                             &                                          &                                        & Yes$^{\,g)}$         &  $(T)$                               &  $(T)$                                 &                   &  \onlinecite{Kirkpatrick_Belitz_Saha_2008a, Kirkpatrick_Belitz_Saha_2008b}                 
\\
\hline\hline\\[-5pt]
\multicolumn{8}{l} {$^{a)}$ Valid for $T\agt T_0$; crossover to exponentially small rates for lower $T$ due to exchange gap.}\\     
\multicolumn{8}{l} {$^{b)}$ Valid for $T\gg T_{\text{ball}}$; crossover to diffusive dynamics for lower $T$.}\\  
\multicolumn{8}{l} {$^{c)}$  Arrows indicate crossover at asymptotically low $T$ due to different dynamics, see Ref.~\onlinecite{Petrova_Tchernyshyov_2011} and the 
                                       discussion in Sec.~\ref{subsec:III.B}.}\\   
\multicolumn{8}{l} {$^{d)}$  Corrections to rates due to ballistic disorder are negative and always small compared to the clean contributions.}\\   
\multicolumn{8}{l} {$^{e)}$ This should properly be interpreted as $T^{1/2}\times T^{1/2}$, as explained in the text.}\\   
\multicolumn{8}{l} {$^{f)}$ Valid for systems on a cubic lattice.}\\
\multicolumn{8}{l} {$^{g)}$  Corrections to rates due to ballistic disorder are positive and always small compared to the clean contributions.}\\  
\end{tabular}
\end{ruledtabular}
\label{table:1}
\end{table*}

\subsection{Theoretical remarks}
\label{subsec:IV.A}

Let us make contact with previous theoretical work on columnar fluctuations in Refs.~\onlinecite{Watanabe_et_al_2014} and \onlinecite{Kirkpatrick_Belitz_2010}. 
In order to relate to the former, we note that the wave-number-resolved quasiparticle rate, which is given by Eq.~(\ref{eq:2.10}) or ({\ref{eq:2.11}) without
the sum over ${\bm k}$, for special directions of ${\bm k}$, scales asymptotically as $T^{5/4}$ in clean systems, with a disorder correction 
that scales as $T$. This behavior does not show in the rate that determines the thermal resistivity, which gets averaged over the Fermi surface. 
Similarly, a wave-number-resolved ``transport rate'' (i.e., the wave-number-resolved single-particle rate with an additional momentum squared 
in the integrand) scales asymptotically as $T^{7/4}$ in clean systems, in agreement with Ref.~\onlinecite{Watanabe_et_al_2014}. Again, this 
behavior does not show in the electrical resistivity, which is determined by the true transport rate that involves an average over the Fermi surface.
Reference~\onlinecite{Kirkpatrick_Belitz_2010} considered intraband scattering due to columnar fluctuations in general, and 
pointed out that in conjunction with ballistic disorder they lead to a $T^{3/2}$ contribution to the
resistivity. While this is generally valid, the particular realization of columnar fluctuations in terms of a superposition of helices proposed
for MnSi in Ref.~\onlinecite{Muehlbauer_et_al_2009} leads to a contribution that is negative and always small compared to the clean $T^3$
contribution, see Sec.~\ref{subsec:III.B} and Table~\ref{table:1}. By contrast, the corresponding interband scattering process considered
in Sec.~\ref{subsec:III.B} comes with a positive prefactor that is much larger, and thus a better candidate for explaining the observations
in MnSi. We will give a more detailed comparison with experimental observations in Sec.~\ref{subsec:IV.B}. 

The last remark raises the question of why ballistic disorder does not always just lead to a small correction to the clean scattering rate.
The answer is that in general it does. However, interband scattering provides a way to avoid this conclusion: Since the clean rate is
exponentially small for $T\alt T_0$, and since disorder eliminates the energy threshold that leads to this suppression, the disorder
``correction'' is actually the leading term in the temperature region $\Tball \alt T \alt T_0$. 

We note that in Dzyaloshinksii-Moriya helimagnets, the helimagnon scattering mechanisms are always suppressed compared
to the skyrmion mechanism, or the FM mechanism at larger wave numbers, due to the small value of $q/\kF$, which in turn is
a consequence of the small spin-orbit coupling. This would be different in systems with a modulated spin order whose wave number
is not small compared to $\kF$. In this context we mention that our general expressions (\ref{eq:2.10}, \ref{eq:2.11}), do apply to
antiferromagnets. In this case $q\approx \kF$, the resonance frequency is linear in $k$, and the relevant susceptibility is given
by the same expression as for HMs, Eq.~(\ref{eq:2.4b}). However, antiferromagnetic magnons are not soft enough to lead to
NFL effects. It is interesting to note, though, that electronic stripe phases have Goldstone modes that have the same form
as in HMs, but with much larger values of $q$.\cite{Kirkpatrick_Belitz_2009a} Such systems would therefore be of interest
to investigate systematically with respect to transport anomalies. 

An important aspect of Eqs.~(\ref{eq:2.10}) and (\ref{eq:2.11}) is that the rate for interband scattering is lacking the gradient-squared
factor that is present in the intraband expression. The reason is that electrons within a given subband cannot couple directly to the
phase of the magnetic order parameter, since that phase has no physical significance. Rather, the coupling is to the gradient of
the phase. For electrons within two different subbands, on the other hand, the coupling is to a phase difference, which does have
a physical meaning.  This was noted before in Ref.~\onlinecite{Bharadwaj_Belitz_Kirkpatrick_2014} in the context of FMs. It is
equally important for the HM cases discussed here, and it is the reason why the clean interband scattering rates have a stronger
temperature dependence than the intraband ones, see Table~\ref{table:1}. 

It may not be obvious why the phase susceptibility in the skyrmionic case, Eq.~(\ref{eq:2.4c}), has qualitatively different forms in the preasymptotic
and asymptotic regions, respectively. The reason is that the static susceptibility, $\chi({\bm k}) = \int (d\omega/\pi) \Im \chi({\bm k},i\Omega\to\omega + i0)$, 
must be equal to the Goldstone mode in both regimes. The latter is known from Ref.~\onlinecite{Ho_et_al_2010}, and the requirement
that both the preasymptotic resonance frequency, Eq.~(\ref{eq:2.3b}), and the asymptotic one, Eq.~(\ref{eq:2.3c}), yield the same
result for the static suscepbility dictates the form of $\chi({\bm k},i\Omega)$ in the second line in Eq.~(\ref{eq:2.4c}).

We finally give an argument for the $3/2$ exponent in the FM case to be exact, rather than a perturbative result that could change at higher
orders in the effective interaction. Consider Eq.~(3.2) in Ref.~\onlinecite{Kirkpatrick_Belitz_2015b}, which gives a general
homogeneity law for the scaling part of the electrical conductivity of a FM. Ballistic disorder eliminates the backscattering factor that
enters the scale dimension of the conductivity, so in our present context the latter is equal to $(d-2) - 2(d-1) = -d$ instead
of $(d-4) - 2(d-1) = -(d+2)$. The relevant dynamical exponent for magnon scattering is $z=2$, which leads to $T^{-3/2}$ for the 
scaling part of $\sigma$ in $d=3$, or $T^{3/2}$ for the scaling part of $\rho$.

\subsection{Comparison with experiments}
\label{subsec:IV.B}

In order to compare our results with experiments, we first need to keep in mind that we ignored all numerical prefactors,
which in any case are model dependent. We therefore can aim only for very rough, order-of-magnitude comparisons
with experimental results. 

We start by discussing the energy scales defined in Sec.~\ref{subsec:II.B} for low-temperature magnets such as MnSi, ZrZn$_2$,
or Ni$_3$Al. The spin-stiffness coefficient is directly measurable, and typically on the order of 
$D \approx 25\,-\,50\,$meV\,\AA$^2$.\cite{Boni_Roessli_Hradil_2011, Bernhoeft_et_al_1982} For $T_1$ this implies
values on the order of 100s of K. The pitch wave number in helimagnets is directly measurable. In MnSi,
$q \approx 0.035$\AA$^{-1}$.\cite{Ishikawa_et_al_1976} This implies that $T_q$ is on the order of 100s of mK. 
For reasonably clean materials ($\rho_0 \alt 1\mu\Omega$cm), the mean-free path is large compared to 
any reasonable value atomic scale, so $\Tball$ will me on the order of 1 mK or less, and can be ignored
in the context of all existing experimental results. This leaves the value of $T_0$, which depends on
the ratio $\lambda/\epsilonF$, which is much harder to determine. From band structure calculations, $\lambda$ 
is typically on the order of thousands of K.\cite{Jeong_Picket_2004, Singh_Mazin_2002, Mazin_Singh_Aguayo_2004} 
This is consistent with Stoner theory, where $\lambda = 6T_1$, although $\lambda$ can be highly anisotropic in 
wave-number space.\cite{Mazin_Singh_Aguayo_2004} If one takes $\lambda = 5,000$K, and $\epsilonF = 10^5K$, 
a typical value for a simple good metal, then one has $\lambda/\epsilonF \approx 0.05$ at most, or $T_0 \approx 1$K. 
However, this is likely misleading for most real materials. The band structures of low-temperature ferromagnets are fairly complicated, 
and the exchange splitting has been suggested to be as large as 0.4 times the relevant effective Fermi energy.\cite{Sigfusson_Bernhoeft_Lonzarich_1984} 
This suggests values of $T_0$ on the order of 10s of K. Based on this discussion, we conclude that reasonable
effective values are, very roughly, $\Tball \approx 1$mK, $T_q \approx 250$mK, $T_0 \approx 10$K, $T_1 \approx 250$K,
$\lambda/\epsilonF \approx 0.1$, and $q/\kF \approx 0.03$, with $T_0$ the most uncertain. 

Given these estimates, the range of validity of Mechanisms 1 and 2 is the same as the temperature range for
which a $T^{3/2}$ behavior, or something close to it, is observed in ZrZn$_2$ and other low-temperature FMs, and
in the disordered phase of the HM MnSi, viz., a few mK to several K. Let us now consider the prefactor of the
$T^{3/2}$ law
\be
\delta\rho(T) = A_{3/2} T^{3/2}\ .
\label{eq:4.1}
\ee
If we ignore all numerical prefactors, as we did in Sec.~\ref{sec:III}, we obtain Eqs.~(\ref{eq:1.2}) and (\ref{eq:1.3})
with $\gamma_1 = \gamma_2 = 1$. With the temperature scales as estimated above, this yields $A_{3/2} \approx 0.001\mu\Omega$cm/K$^{3/2}$,
which is a factor of 10 smaller than what is observed in ZrZn$_2$, and a factor of 100 smaller than what is observed
in MnSi. If we take the numerical factors seriously, we have, after various cancellations, a factor of $\pi$ from the 
spectrum of the susceptibility, and a factor of about 3 from the dimensional integral in Eq.~(\ref{eq:3.3a}), or a 
factor of about 6 from the one that enters Eq.~(\ref{eq:3.5}). This leads to $\gamma_1 \approx 10$, and $\gamma_2 \approx 20$.
While these estimates should not be taken too seriously, they indicate that Mechanism 1 produces a prefactor $A_{3/2}$ that
is very close to what is observed in ZrZn$_2$, and Mechanism 2 produces a $A_{3/2}$ that is within a factor of
5 of what is observed in the disordered phase of MnSi. For the latter, we assume that columnar fluctuations exist in this phase, 
as was proposed in Ref.~\onlinecite{Kirkpatrick_Belitz_2010}. There is strong experimental evidence,\cite{Pfleiderer_et_al_2004} 
as well as theoretical arguments,\cite{Tewari_Belitz_Kirkpatrick_2006} for this phase to be a strongly correlated liquid-like phase.
The associated strong fluctuations are expected to enhance the prefactor compared to our simple relaxation-rate
considerations and may well account for the remaining factor of 5. 

Let us now briefly discuss the ordered phase of MnSi, where the observed resistivity behavior is $T^2$, see Fig.~\ref{fig:2}.
In this context it is important to remember that there are many contributions to the electron scattering rate, e.g., due to the
Coulomb interaction, phonons, and other excitations, that lead to contributions that go as $T^s$ with $s\geq 2$.
We have focused on the contributions due to magnetic Goldstone modes, which have the remarkable property that they
lead to a $T^{3/2}$ behavior. In addition to whatever scattering mechanism is producing the $T^2$ behavior in the helically
ordered phase of MnSi, we therefore expect a $T^{3/2}$ contribution for $T>T_q$,
where MnSi effectively behaves like a FM and Mechanism 1 applies, and a $T \ln T$ contribution for $T<T_q$, where Eq.~(\ref{eq:3.10}) applies
(note, however, the caveats mentioned after Eq.~(\ref{eq:3.10}), which requires a thorougher investigation).
Consider the former regime, and write the resistivity as
\bse
\label{eqs:4.2}
\be
\delta\rho = A_{3/2} T^{3/2} + A_2 T^2 \quad (T\agt T_q)\ .
\label{eq:4.2a}
\ee
For $A_2$ we take the observed value $A_2 \approx 0.03\mu\Omega$cm/K$^{2}$.\cite{Pfleiderer_2007} For $A_{3/2}$ we expect
a value comparable to that observed in ZrZn$_2$ and other low-temperature FMs, $A_{3/2} \approx 0.01\mu\Omega$cm/K$^{3/2}$. The $T^{3/2}$
term will then dominate for $T \alt (A_{3/2}/A_2)^2 \approx 100$mK, which is less than $T_q$. The $T^{3/2}$ contribution will therefore
be a small correction to the dominant $T^2$ behavior at all temperature where Eq.~(\ref{eq:3.3b}), or equivalently Eq.~(\ref{eq:1.2}), applies
to MnSi. In the latter regime, the helical nature of the Goldstone modes will be
apparent, and various nonanalytic contributions to $\delta\rho$ must be present. The $T^{5/2}$ contribution from clean intraband
HM scattering is too weak to be observable, and the disorder contribution due to interband scattering comes with the caveats
mentioned after Eq.~(\ref{eq:3.10}) and requires a more detailed investigation. Ignoring these caveats, and ignoring the logarithm, we expect
\be
\delta\rho = A_1 T + A_2 T^2 \quad (T \alt T_q)\ .
\label{eq:4.2b}
\ee
\ese
With the same parameters as used above one finds $A_1 \approx 10^{-3}\mu\Omega$cm/K. The $T\ln T$ term will thus dominate over the $T^2$ term
only for temperatures small compared to about 30mK, although the logarithm, and a more thorough determination of the prefactor,
might change this estimate. We conclude that any nonanalytic contribution
to $\delta\rho$ in the helically ordered phase would require a precision experiment at temperatures smaller than at least 100mK. 

The $T^s$ behavior with $1.5 \alt s < 2$ observed in ZrZn$_2$ is most likely the result of a $T^{3/2}$ behavior due to Mechanism 1
in addition to a $T^2$ contribution from other scattering mechanisms. If one takes Eq.~(\ref{eq:4.2a}) with $A_{3/2}/A_2 \approx 1$K$^{1/2}$,
then between about 100mK and 10K the behavior is well represented by a single $T^{1.7}$ law. Another question is why the transport
anomaly is observed in the magnetically disordered phase as well as in the ordered one. A plausible answer lies in the fact that the
magnetic transition is first order. As has been discussed in Ref.~\onlinecite{Kirkpatrick_Belitz_2016b}, even very weak quenched
disorder leads to static islands or droplets of the ordered phase within the disordered one, which explains the commonly observed
phase separation in the vicinity of first-order transitions (see Refs.~\onlinecite{Uemura_et_al_2007, Pfleiderer_et_al_2010, Gat-Malureanu_et_al_2011}
for examples). It also explains why the Goldstone modes of the ordered phase, i.e., the FM magnons, can still contribute to
the scattering of electrons even in the magnetically disordered phase. 

We have focused on MnSi and ZrZn$_2$ in this discussion since these are the experimentally
best studied systems.
There are, however, many other examples of low-temperature ferromagnets where NFL behavior has been observed:
Off-stoichiometric NiGa$_3$ is a FM for Ni-rich concentrations and displays a
NFL $T^{3/2}$ behavior with a coefficient of $A_{3/2}\approx 0.04\mu\Omega$cm/K$^{3/2}$.\cite{Fluitman_et_al_1973, Pfleiderer_2007}
For smaller Ni concentrations the material is paramagnetic and $\delta\rho$ is Fermi-liquid-like. The stoichiometric compound
has the smallest residual resistivity, $\rho_0 \approx 1\mu\Omega$cm. In Ni$_3$Al under pressure, for samples with $\rho_{0}\approx 1\mu\Omega$cm, 
a $T^{3/2}$ dependence, or a power law close to that, is observed on either side of the ferromagnetic transition with a coefficient 
$A_{3/2}\approx 0.02\mu\Omega$cm/K$^{3/2}$, see Ref.~\onlinecite{Niklowitz_et_al_2005}.
Similar observations in more disordered systems include
Nb$_{1-x}$Fe$_{2+x}$, for which a $T^{3/2}$  behavior has been observed on the non-magnetic side of a magnetic 
transition.\cite{Brando_et_al_2008} In this case, $\rho_{0}\approx 5\mu\Omega$cm. In (Ni$_{1-x}$Pd$_x$)$_3$Al, with 
$\rho_0 \approx 10\mu\Omega$cm, a $T^{3/2}$ behavior was found on both sides of the FM transition.\cite{Sato_1975}
All of these observations can be understood at a semi-quantitative level by a discussion analogous to the one given
above, which slightly different parameter values. 

Finally, our results suggest a number of possible experiments. One interesting question is whether or not there is a small NFL contribution 
to the resistivity in the ordered phase of MnSi. As discussed above, this would require a precise determination of the resistivity at very
low temperatures, probably lower than 100mK. This would entail subtracting the $T^2$ contribution to reveal any underlying NFL
behavior. Of particular interest is the logarithmic dependence on the sample size predicted by Eq.~(\ref{eq:3.10}). It suggests
that, in systems small enough for the spin-orbit coupling not to intervene, the resistivity ({\em not} just the resistance) will change 
with changing sample size, which reflects the soft helimagnon excitations in a HM phase. Additional theoretical investigations are 
also called for to make this prediction more precise.\cite{spin-orbit_footnote} Another interesting question is the magnetic-field dependence
of the NFL resistivity. For the FM mechanism, Eq.~(\ref{eq:1.2}), the theory predicts that in a field $H$ such that
$\mu_{\text B} H > T$ the magnon-induced resistivity contribution becomes $\delta\rho_{\text{FM}} \propto T^2/H^{1/2}$. A study
of the magnetoresistance of ZrZn$_2$ would be very interesting in that respect.

\acknowledgements
This work was supported by the NSF under Grant Nos. DMR-1401449 and DMR-1401410. This work was initiated at the 
Aspen Center for Physics, supported by the NSF under Grant No. PHYS-1066293, and continued at the Telluride Science
Research Center (TSRC).

\appendix*
\section{Disordered helimagnets in the ballistic limit}
\label{app:A}

In this appendix we discuss the relation between the simple arguments given in Sec.~\ref{subsubsec:II.C.3} about the
effects of disorder in the ballistic regime and a detailed calculation. This also sheds light on the confusing issue of the
sign of the disorder correction to the relaxation rates, which is different for the case of helimagnon interband scattering
from all of the other cases.

To make the salient point it suffices to consider the single-particle rate, which is given as the imaginary part of the
electronic self energy. In the presence of weak disorder, the leading correction to the clean result is given by the
diagram shown in Fig.~\ref{fig:5}. Analytically, this contribution is (see Eq.~(3.11) in Ref.~\onlinecite{Kirkpatrick_Belitz_Saha_2008b})
\begin{align}
\frac{1}{\tau_{\text{sp}}} &\propto \frac{1}{\tau}\,\frac{D \epsilonF^2 q^2}{\NF}\, \frac{1}{V}\sum_{\bm k} \text{Im} L^{++,-}({\bm k}) 
\nonumber\\
& \quad \times \int_{-\infty}^{\infty} \frac{du}{\pi}\, n_{\text{F}}(u/T)\,\chi''({\bm k},u)\ .
\tag{A.1a}
\label{eq:A.1a}
\end{align}
Here $n_{\text{F}} = 1/(e^x+1)$ is the Fermi distribution function, and 
\be
L^{++,-}({\bm k}) = \frac{1}{V}\sum_{\bm p} \gamma({\bm k},{\bm p})\,\gamma({\bm k},{\bm p}-{\bm k})\ G_R^2({\bm p})\,G_A({\bm p}-{\bm k})\ ,
\tag{A.1b}
\label{eq:A1b}
\ee
and we have kept only terms that contribute to the leading temperature dependence of $1/\tau_{\text{sp}}$.
$G_R$ and $G_A$ are the retarded and advanced Green function, respectively. In the case of intraband scattering
all three Green functions belong to the same Fermi surface. The couplings $\gamma$ depend in general on both the
transferred momentum and the momentum of the incoming or outgoing electron. In the case of interband scattering, they
are constants. In the case of interband scattering, they are gradients that in Eq.~({\ref{eq:2.11}) we have represented by
the transferred momentum, $\gamma({\bm k},{\bm p}) \propto {\bm k}$. This is qualitatively correct in the case of 
skyrmionic fluctuations, where $k_z$ provides the leading temperature scaling, see Eqs.~(\ref{eqs:2.3}), and the
angular part of the integral in Eq.~(\ref{eq:A.1a}) is
\be
\int_{-1}^{1} d\eta\ \frac{1}{(\vF k \eta - i0)^2} = \frac{-2}{(\vF k)^2}\ .
\tag{A.2a}
\label{eq:A.2a}
\ee
However, in the case of helimagnon scattering, where the leading scaling is provided by ${\bm k}_{\perp}$, the dependence of
$\gamma$ on ${\bm k}_{\perp}$ comes in the form $({\bm k}_{\perp}\cdot{\bm p}_{\perp})p_z$, see Eq.~(2.18c) in
Ref.~\onlinecite{Kirkpatrick_Belitz_Saha_2008a}. This introduces an angular dependence of the integrand that is not
present in the schematic representation in Eq.~(\ref{eq:2.11}). The relevant angular integral then is
\be
\int_0^{2\pi} d\varphi\ \frac{\cos^2\varphi}{(\vF k_{\perp}\cos\varphi - i0)^2} = \frac{1}{(\vF k_{\perp})^2}\ .
\tag{A.2b}
\label{eq:A.2b}
\ee
The sign of the disorder correction is thus different in the two cases. For the case of helimagnon interband scattering, where Eq.~(\ref{eq:A.2b}) applies, we recover the
result of Ref.~\onlinecite{Kirkpatrick_Belitz_Saha_2008a}, viz.
\be
\delta\left({1/\tau_{\text{sp}}}\right) \propto \frac{1}{\tau}\,\left(\frac{q}{\kF}\right)^5 \frac{\epsilonF}{\lambda} \left[\frac{-\Lambda}{T_q} + \ln 2\,\frac{T}{T_q}\right]\ .
\label{eq:A.3}
\tag{A.3}
\ee
Here $\Lambda$ is a UV energy cutoff. The cutoff-dependent constant contribution to the rate is negative, so the effect is antilocalizing,
and the universal temperature-dependent contribution is accordingly positive. For the interband skyrmion case, on the other hand,
Eq.~(\ref{eq:A.2a}) applies. The effect then has the opposite sign and is localizing, i.e., the temperature-dependent disorder correction
to the rate is negative. 

Now compare these results with the simple argument given in Sec.~\ref{subsubsec:II.C.3}, which replaces $\text{Im} L^{++,-}({\bm k})$ with
\be
L({\bm k}) = \int_{-1}^{1} d\eta\,\frac{1/\tau}{(\vF k \eta)^2 + 1/\tau^2} = \frac{1}{\vF k} \left[\frac{\pi}{2} - \frac{1}{\vF k \tau} + \ldots\right]\ .
\tag{A.4}
\label{eq:A.4}
\ee
The simple argument thus yields the correct momentum or temperature scaling in both cases, but produces the wrong sign in the
helimagnon case. 

Finally, consider the case of interband scattering, where $G_R$ belong to one subband, and $G_A$ to the other. The relevant angular
integral then is
\be
\int_{-1}^{1} d\eta\,\frac{1}{(\vF k \eta - 2\lambda - i0)^2} = \frac{1}{2\lambda^2}
\tag{A.5a}
\label{eq:A.5a}
\ee
for $k\to 0$. The simple smearing argument replaces this by
\begin{align}
\int_{-1}^{1} d\eta\,\frac{1}{(\vF k \eta - 2\lambda)^2 + 1/\tau^2} &= \frac{1}{2\lambda^2} \quad 
\nonumber\\
& \hskip -50pt (\vF k \ll \lambda\ ,\ \lambda \gg 1/\tau)\ ,
\tag{A.5b}
\label{eq:A.5b}
\end{align}
and thus produces the correct result. 

In conclusion, the simple $\delta$-function smearing argument from Sec.~\ref{subsubsec:II.C.3} yields the correct temperature scaling
for all of the cases considered, and the correct sign of the disorder correction except in the case of intraband helimagnon scattering.


\begin{thebibliography}{54}
\bibitem[{mic()}]{microscopic_scales_footnote}
\bibinfo{note}{Depending on the material, the microscopic energy scale could be
  a band width rather than the Fermi energy.}
\end{thebibliography}

\end{document}